\documentclass[conference,a4paper]{IEEEtran}
\usepackage{graphicx}
\usepackage{subfigure}
\usepackage{pgf}
\usepackage{tikz}
\usepackage{pgfplots}
\usetikzlibrary{arrows,automata}
\usetikzlibrary{positioning}
\usetikzlibrary{backgrounds}
\usetikzlibrary{matrix}
\usetikzlibrary{decorations.pathmorphing}
\usetikzlibrary{fit}
\usetikzlibrary{calc}

\begin{document}

\newtheorem{lemm}{Lemma}
\newtheorem{theo}[lemm]{Theorem}
\newtheorem{prop}[lemm]{Proposition}
\newtheorem{defi}[lemm]{Definition}
\newtheorem{coro}[lemm]{Corollary}
\newtheorem{exam}{Example}
\renewcommand{\IEEEQED}{\IEEEQEDopen}

\sloppy

\title{Non-homogeneous Two-Rack Model for \\Distributed Storage Systems}

\author{
  \IEEEauthorblockN{
    Jaume Pernas\IEEEauthorrefmark{1},
    Chau Yuen\IEEEauthorrefmark{1},
    Bernat Gast\'on\IEEEauthorrefmark{2} and
    Jaume Pujol\IEEEauthorrefmark{2}}
  \IEEEauthorblockA{
    \IEEEauthorrefmark{1}Singapore University of Technology and
    Design,
    \IEEEauthorrefmark{2}Universitat Autonoma de Barcelona\\
    Email: \{jaume\_pernas,yuenchau\}@sutd.edu.sg, \{bernat.gaston,jaume.pujol\}@uab.cat}

}




\maketitle

\begin{abstract}
In the traditional two-rack distributed storage system (DSS) model,
due to the assumption that the storage capacity of each node is the
same, the minimum bandwidth regenerating (MBR) point becomes
infeasible. In this paper, we design a new non-homogeneous two-rack
model by proposing a generalization of the threshold function used 
to compute the tradeoff curve. We prove that by having the nodes 
in the rack with higher regenerating bandwidth stores more information, 
all the points on the tradeoff curve, including the MBR point, become 
feasible. Finally, we show how the non-homogeneous two-rack model 
outperforms the traditional model in the tradeoff curve between 
the storage per node and the repair bandwidth.
\end{abstract}

\section{Introduction}

Cloud storage has been consolidated as a growing paradigm, as it provides a convenience solution for online storage that
is accessible with any device at anywhere and anytime.

To ensure reliability, in practice, cloud storage is implemented in
terms of distributed storage system (DSS), where several
geographically distributed storage nodes collaboratively to provide
storage or backup services. Such distributed system provides
diversity and achieves fault-tolerance against catastrophic failure,
it also minimizes the probability of losing the stored data and
maximizes the data availability.

Erasure coding has been proven in \cite{rod05,wea02} as an effective
technique for such DSS. Through the use of erasure coding, fault
tolerance level is improved and the size of stored data is
minimized. Moreover, \cite{dim10} shows that with the use of
regenerating codes, not only achieves most of the improvements of
erasure coding, but also minimizes the amount of data needed to
regenerate a failed node. Since then, the theoretical and
fundamental tradeoffs among the system resources, e.g. storage
capacity and repair bandwidth, has been discovered. Several novel
coding schemes, e.g. \cite{ras11,rou10}, have been constructed to
achieve the tradeoff curve in certain special points, e.g. minimum
storage regenerating (MSR) and minimum bandwidth regenerating (MBR).

The previous theoretical results were assuming a symmetric and
homogeneous model in terms of data storage and repair
bandwidth. However, in a realistic implementation, not all nodes are
equal in terms of storage size, repair bandwidth, or even
reliability. By considering the difference in terms of repair
bandwidth, \cite{Akh10} proposes a DSS model where there is a static
classification of storage nodes based on their repair
bandwidth, storage nodes are divided into two groups, one is ``cheap
bandwidth'' and another ``expensive bandwidth''.

To generalize the above static model, \cite{Ber13} considers that
the storage nodes are organized in two racks. The repair
bandwidth cost between nodes within the same rack is much lower than between nodes across different racks. This situation introduces a dynamic model, where the
classification of ``cheap/expensive bandwidth'' falls on the
relation between two nodes. The bandwidth between two nodes is
``cheap'' if both are from the same rack and ``expensive''
otherwise. Using this two-rack model, the authors in \cite{Ber13}
have shown the tradeoff between bandwidth and storage with repair
cost. In this paper, our focus is on such two-rack model due to its
practical implication, for example, consider a DSS that spans across
two countries, it can be easily modeled with two-rack model where
the storage nodes within the same country enjoy ``cheap bandwidth'',
while the storage nodes across different countries have ``expensive
bandwidth''. Unfortunately, the authors in \cite{Ber13} show that it
is infeasible to achieve the MBR point for such two-rack model.

While the previous models, e.g. the static model in \cite{Akh10} and
the two-rack model in \cite{Ber13}, have considered a DSS with
different repair bandwidth among the storage nodes, all of
them assume the storage nodes have the same storage capacity. Recent
development have included the emergence of non-homogeneous DSS that
pool together nodes with truly different characteristics, including
the storage size. The capacity of such non-homogeneous DSS with
different storage size and repair bandwidth has been studied
in \cite{Ern12}. Coding scheme for a non-homogeneous storage system
with one super-node that is more reliable and has more storage
capacity is studied in \cite{Van12}.

In this paper, we show that by considering a non-homogeneous model,
where all the nodes have different storage size and repair
bandwidth, not only such model is closer to practical system, it
also provides a solution to the problem of infeasible MBR point in
the two-rack model mentioned above. We design a two racks DSS such
that storage size at each node is depending on the repair
bandwidth of each rack, and prove that such design can achieve the
MBR point.

Our paper is organized as follows. In
Section~\ref{sec:ModelsDescription} we describe various DSS models.
We start with the symmetric model used to explain the information
flow graph. Then, we explain the static cost model because it is the
first model presenting storage nodes with different repair
bandwidth. Then, we introduce the two-rack model as a generalization
of the static cost model. We start Section~\ref{sec::results} by
presenting the problem of the two-rack model on infeasible MBR
point, and then propose our solution of creating a non-homogeneous
two-rack model. Finally, we conclude the paper in
Section~\ref{sec::conc}.

\section{Previous Models of DSS}
\label{sec:ModelsDescription}

\begin{figure*}[ht!]
     \begin{center}
        \subfigure[Symmetric model ($\gamma = 3\beta$).]{%
            \label{mod::a}             
            \begin{tikzpicture}[shorten >=1pt,->]
              \tikzstyle{vertix}=[circle,fill=black!25,minimum size=18pt,inner sep=0pt, node distance = 0.8cm,font=\tiny]
              \tikzstyle{invi}=[circle]
              \tikzstyle{background}=[rectangle, fill=gray!10, inner sep=0.2cm,rounded corners=5mm]

              \node[vertix]  (vin_1)  {$v_{in}^1$};
              \node[vertix, below of=vin_1,] (vin_2)  {$v_{in}^2$};
              \node[vertix, below of=vin_2] (vin_3)  {$v_{in}^3$};
              \node[vertix, below of=vin_3] (vin_4)  {$v_{in}^4$};

              \node[vertix, right of=vin_1] (vout_1)  {$v_{out}^1$};
              \node[vertix, right of=vin_2] (vout_2)  {$v_{out}^2$};
              \node[vertix, right of=vin_3] (vout_3)  {$v_{out}^3$};
              \node[vertix, right of=vin_4] (vout_4)  {$v_{out}^4$};

            \draw[-,color=red,thick] (-0.2,0.2) -- (1,-0.2);
            \draw[-,color=red,thick] (-0.2,-0.2) -- (1,0.2);

              \node[vertix, right of=vout_3, node distance = 1.5cm] (vin_5)
            {\scriptsize{$v_{in}^5$}};
              \node[vertix, right of=vin_5] (vout_5)
            {\scriptsize{$v_{out}^5$}};

              \path[->, bend left=15,font=\footnotesize] (vout_2) edge
            node[anchor=south,above]{$\beta$} (vin_5);
              \path[->, bend left=10,font=\footnotesize] (vout_3) edge
            node[anchor=south,above]{$\beta$}(vin_5);
              \path[->, bend left=20,font=\footnotesize] (vout_4) edge
            node[anchor=south,above]{$\beta$}(vin_5);

             \foreach \from/\to in {1,2,3,4,5}
              { \path[->,font=\footnotesize] (vin_\from) edge node[anchor=south] {$\alpha$}
            (vout_\to); }
             \end{tikzpicture}
        }\qquad\qquad
        \subfigure[Static cost model ($\gamma = 2\beta_c + 2\beta_e$).]{%
           \label{mod::b}
\begin{tikzpicture}[shorten >=1pt,->]
  \tikzstyle{vertix}=[circle,fill=black!25,minimum size=18pt,inner sep=0pt,
node distance = 0.8cm,font=\tiny]
  \tikzstyle{vertixfast}=[circle,fill=blue!25,minimum size=18pt,inner sep=0pt, node distance = 0.8cm,font=\tiny]
  \tikzstyle{invi}=[circle]
  \tikzstyle{background}=[rectangle, fill=gray!10, inner sep=0.2cm,rounded
corners=5mm]


  \node[vertix]  (vin_1)  {$v_{in}^1$};
  \node[vertix, below of=vin_1] (vin_2)  {$v_{in}^2$};
  \node[vertix, below of=vin_2] (vin_3)  {$v_{in}^3$};
  \node[vertixfast, below of=vin_3] (vin_4)  {$v_{in}^4$};
  \node[vertixfast, below of=vin_4] (vin_5)  {$v_{in}^5$};
  \node[vertixfast, below of=vin_5] (vin_6)  {$v_{in}^6$};

  \node[vertix, right of=vin_1] (vout_1)  {$v_{out}^1$};
  \node[vertix, right of=vin_2] (vout_2)  {$v_{out}^2$};
  \node[vertix, right of=vin_3] (vout_3)  {$v_{out}^3$};
  \node[vertixfast, right of=vin_4] (vout_4)  {$v_{out}^4$};
  \node[vertixfast, right of=vin_5] (vout_5)  {$v_{out}^5$};
  \node[vertixfast, right of=vin_6] (vout_6)  {$v_{out}^6$};

\draw[-,color=red,thick] (-0.2,0.2) -- (1,-0.2);
\draw[-,color=red,thick] (-0.2,-0.2) -- (1,0.2);

\draw[-,color=red,thick] (-0.2,-3.8) -- (1,-4.2);
\draw[-,color=red,thick] (-0.2,-4.2) -- (1,-3.8);

  \node[vertix, right of=vout_3, node distance = 1.5cm] (vin_7)
{\scriptsize{$v_{in}^7$}};
  \node[vertix, right of=vin_7] (vout_7)
{\scriptsize{$v_{out}^7$}};
  \node[vertixfast, right of=vout_6, node distance = 1.5cm] (vin_8)
{\scriptsize{$v_{in}^8$}};
  \node[vertixfast, right of=vin_8] (vout_8)
{\scriptsize{$v_{out}^8$}};

  \path[->, bend left=15,font=\footnotesize] (vout_2) edge
node[anchor=south,above]{$\beta_e$} (vin_7);
  \path[->, bend left=10,font=\footnotesize] (vout_3) edge
node[anchor=south,above]{$\beta_e$}(vin_7);
  \path[->, bend left=20,font=\footnotesize] (vout_4) edge
node[anchor=south,above]{$\beta_c$}(vin_7);
  \path[->, bend left=20,font=\footnotesize] (vout_5) edge
node[anchor=south,above]{$\beta_c$}(vin_7);

  \path[->, bend left=15,font=\footnotesize] (vout_2) edge
node[anchor=south,above]{$\beta_e$} (vin_8);
  \path[->, bend left=10,font=\footnotesize] (vout_3) edge
node[anchor=south,above]{$\beta_e$}(vin_8);
  \path[->, bend left=20,font=\footnotesize] (vout_4) edge
node[anchor=south,above]{$\beta_c$}(vin_8);
  \path[->, bend left=20,font=\footnotesize] (vout_5) edge
node[anchor=south,above]{$\beta_c$}(vin_8);

 \foreach \from/\to in {1,2,3,4,5,6,7,8}
  { \path[->,font=\footnotesize] (vin_\from) edge node[anchor=south] {$\alpha$}
(vout_\to); }

 \end{tikzpicture}
        }\qquad\qquad
        \subfigure[Traditional two-rack model ($\gamma^1 = \beta_c + 2\beta_e$, $\gamma^2 = 2\beta_c + \beta_e$).]{%
            \label{mod::c}
\begin{tikzpicture}[shorten >=1pt,->]
  \tikzstyle{vertix}=[circle,fill=black!25,minimum size=18pt,inner sep=0pt,
node distance = 0.8cm,font=\tiny]
  \tikzstyle{invi}=[circle]
  \tikzstyle{background}=[rectangle, fill=gray!10, inner sep=0.2cm,rounded
corners=5mm]


  \node[vertix]  (vin_1)  {$v_{in}^1$};
  \node[vertix, below of=vin_1,] (vin_2)  {$v_{in}^2$};
  \node[vertix, below of=vin_2, node distance = 1.5cm] (vin_3)  {$v_{in}^3$};
  \node[vertix, below of=vin_3] (vin_4)  {$v_{in}^4$};
  \node[vertix, below of=vin_4] (vin_5)  {$v_{in}^5$};
  \node[vertix, below of=vin_5] (vin_6)  {$v_{in}^6$};

  \node[vertix, right of=vin_1] (vout_1)  {$v_{out}^1$};
  \node[vertix, right of=vin_2] (vout_2)  {$v_{out}^2$};
  \node[vertix, right of=vin_3] (vout_3)  {$v_{out}^3$};
  \node[vertix, right of=vin_4] (vout_4)  {$v_{out}^4$};
  \node[vertix, right of=vin_5] (vout_5)  {$v_{out}^5$};
  \node[vertix, right of=vin_6] (vout_6)  {$v_{out}^6$};

\draw[-,color=red,thick] (-0.2,0.2) -- (1,-0.2);
\draw[-,color=red,thick] (-0.2,-0.2) -- (1,0.2);
\draw[-,color=red,thick] (-0.2,-2.1) -- (1,-2.5);
\draw[-,color=red,thick] (-0.2,-2.5) -- (1,-2.1);

  \node[vertix, right of=vout_2, node distance = 1.5cm] (vin_7)
{\scriptsize{$v_{in}^7$}};
  \node[vertix, right of=vin_7] (vout_7)
{\scriptsize{$v_{out}^7$}};

  \node[vertix, right of=vout_3, node distance = 1.5cm] (vin_8)
{\scriptsize{$v_{in}^8$}};
  \node[vertix, right of=vin_8] (vout_8)
{\scriptsize{$v_{out}^8$}};

  \path[->, bend left=15,font=\footnotesize] (vout_2) edge
node[anchor=south,above]{$\beta_c$} (vin_7);
  \path[->, bend left=10,font=\footnotesize] (vout_4) edge
node[anchor=south,above]{}(vin_7);
  \path[->, bend left=20,font=\footnotesize] (vout_5) edge
node[anchor=south,above]{$2\beta_e$}(vin_7);

  \path[->, bend left=15,font=\footnotesize] (vout_2) edge
node[anchor=south,above]{$\beta_e$} (vin_8);
  \path[->, bend left=10,font=\footnotesize] (vout_4) edge
node[anchor=south,above]{$\beta_c$}(vin_8);
  \path[->, bend left=10,font=\footnotesize] (vout_5) edge
node[anchor=south,above]{$\beta_c$}(vin_8);

 \foreach \from/\to in {1,2,3,4,5,6,7,8}
  { \path[->,font=\footnotesize] (vin_\from) edge node[anchor=south] {$\alpha$}
(vout_\to); }

 \end{tikzpicture}
        }%
    \end{center}
    \caption{%
        Different models of DSS.
     }%
   \label{fig::models}
\end{figure*}
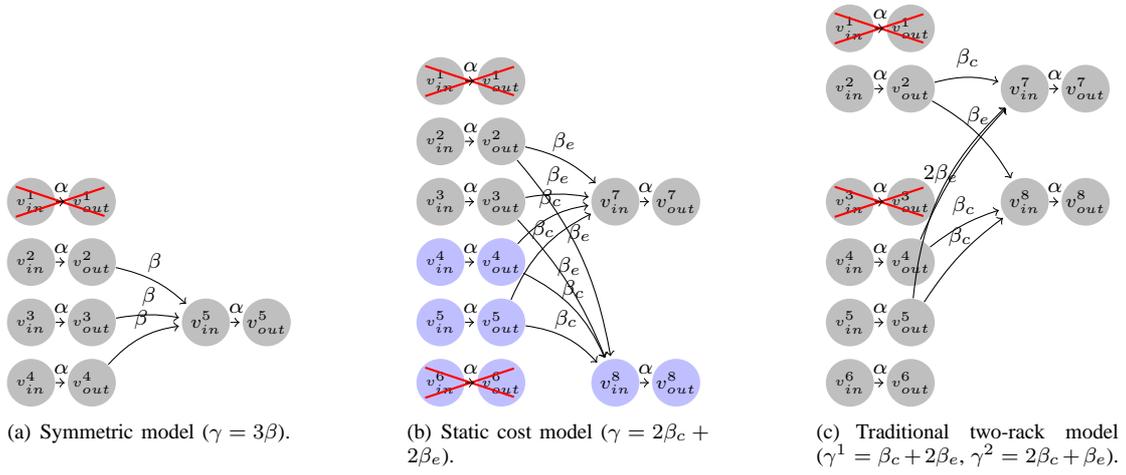

In this section, we present three different models of DSS. In
Subsection~\ref{sub:first_model} we show the 
symmetric model, where the repair bandwidth and the storage
size is the same for all the nodes. We will use this model to
explain the information flow graph, which is essential for the
readers to better understand our contribution at a later time. In
Subsection~\ref{sub:static_cost_model}, we present a static cost DSS
model, where the nodes are divided into two groups, namely cheap and
expensive, based on their repair bandwidth. In this case,
since the nodes are always cheap or expensive, no matter who is
connecting to them, the repair bandwidth is always the same.
This static cost model is a particular case of the two-rack model
that will be presented in Subsection~\ref{sub:two_racks_model}. In
the two-rack model, the cheap or expensive connection depends on the
helper nodes and the newcomer. Hence, there are two different
repair bandwidths. Figure~\ref{fig::models} shows the
differences between the three models. We will discuss each model in
great details, as understanding them is the key to understand our
contribution.

\subsection{Symmetric Model} 
\label{sub:first_model}

In~\cite{dim10}, Dimakis et. al. first introduced a symmetric
distributed storage model, where every storage node has the same
storage size and the same repair bandwidth. As such the
repair cost for every storage node is the same. Moreover, the
fundamental tradeoff between the amount of stored data per node and
the repair bandwidth can be obtained by analyzing the mincut of the
information flow graph.

The information flow is a directed acyclic graph including three types of nodes: $(i)$ A single source node (S), $(ii)$ Some intermediate nodes and $(iii)$ Data collectors (DC). The source node is the source of original data file, intermediate nodes are storage nodes and each data collector corresponds to a request to reconstructing the original file. Each storage node is represented by pairs of incoming and outgoing nodes connected by a directional edge whose capacity is the corresponding storage capacity $\alpha$ of this storage node. Moreover, it is assumed edges departing the storage nodes and arriving to a DC node have an infinite capacity. This reflects the fact that DC nodes have access to all stored data of the surviving nodes they are connected to.

\begin{figure}
\centering
\begin{tikzpicture}[shorten >=1pt,->]
  \tikzstyle{vertix}=[circle,fill=black!25,minimum size=18pt,inner sep=0pt,
node distance = 0.8cm,font=\tiny]
  \tikzstyle{invi}=[circle]
  \tikzstyle{background}=[rectangle, fill=gray!10, inner sep=0.2cm,rounded
corners=5mm]

  \node[vertix] (s) {$S$};

  \node[vertix, right=1cm of s] (vin_2)  {$v_{in}^2$};
  \node[vertix, below of=vin_2] (vin_3)  {$v_{in}^3$};
  \node[vertix, above of=vin_2] (vin_1)  {$v_{in}^1$};
  \node[vertix, below of=vin_3] (vin_4)  {$v_{in}^4$};

 \foreach \to in {1,2}
   {\path (s) edge[bend left=20,font=\tiny] node[anchor=south,above]{$\infty$}
(vin_\to);}
 \foreach \to in {3,4}
   {\path (s) edge[bend right=20,font=\tiny]  node[anchor=south,above]{$\infty$}
 (vin_\to);}

  \node[vertix, right of=vin_1] (vout_1)  {$v_{out}^1$};
  \node[vertix, right of=vin_2] (vout_2)  {$v_{out}^2$};
  \node[vertix, right of=vin_3] (vout_3)  {$v_{out}^3$};
  \node[vertix, right of=vin_4] (vout_4)  {$v_{out}^4$};

  \node[vertix, right of=vout_4, node distance = 1.5cm] (vin_5)
{\scriptsize{$v_{in}^5$}};
  \node[vertix, right of=vin_5] (vout_5)
{\scriptsize{$v_{out}^5$}};

  \node[vertix, right of=vout_1, above of=vin_5, node distance = 1.5cm] (vin_6)
{\scriptsize{$v_{in}^6$}};
  \node[vertix, right of=vin_6] (vout_6)
{\scriptsize{$v_{out}^6$}};

  \node[vertix, right of=vout_6, node distance = 1.5cm] (DC)
{DC};

   \path (vout_5) edge[bend right=20,font=\tiny]
node[anchor=south,above]{$\infty$}
 (DC);
   \path (vout_6) edge[bend left=10,font=\tiny]
node[anchor=south,above]{$\infty$}
 (DC);

  \path[->, bend left=15,font=\tiny] (vout_2) edge
node[anchor=south,above]{$\beta$} (vin_5);
  \path[->, bend left=10,font=\tiny] (vout_3) edge
node[anchor=south,above]{$\beta$}(vin_5);
  \path[->, bend left=20,font=\tiny] (vout_1) edge
node[anchor=south,above]{$\beta$}(vin_5);

  \path[->, bend left=15,font=\tiny] (vout_1) edge
node[anchor=south,above]{$\beta$} (vin_6);
  \path[->, bend left=10,font=\tiny] (vout_2) edge
node[anchor=south,above]{$\beta$}(vin_6);
  \path[->, bend left=10,font=\tiny] (vout_5) edge
node[anchor=south,above]{$\beta$}(vin_6);

 \foreach \from/\to in {1,2,3,4,5,6}
  { \path[->,font=\tiny] (vin_\from) edge node[anchor=south] {$\alpha$}
(vout_\to); }

  \draw[-,color=red,thick] (1.2,-0.5) -- (2.7,-1.2);
  \draw[-,color=red,thick] (1.2,-1.2) -- (2.7,-0.5);

  \draw[-,color=red,thick] (1.2,-1.2) -- (2.7,-1.9);
  \draw[-,color=red,thick] (1.2,-1.9) -- (2.7,-1.2);
\end{tikzpicture}

\caption{Information flow graph corresponding to a $[4,2,3]$ regenerating code.
}
\label{fig::modelA}
\end{figure}
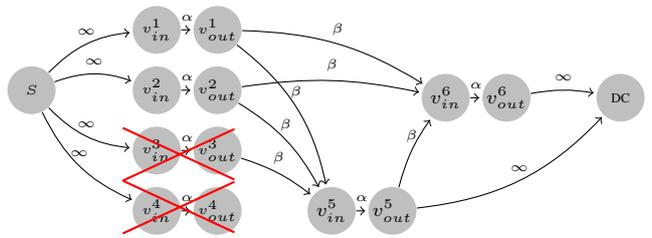

The graph evolves constantly across time to capture any changes happening throughout the network. This graph starts from the source node. It is the only active node at the first step. The total number of storage nodes is $n$ and the source node divides the original data file of size $M$ into $k$ pieces. These $k$ pieces are encoded to $n$ data fragments each to be stored in one of existing storage nodes through direct edges of infinite capacity. In the case that a storage node leaves the system or a failure occurs, this node is replaced by a new one, called the newcomer node. The newcomer connects to $d$ active nodes out of $n - 1$ existing nodes and downloads $\beta$ bits from each. Accordingly, the corresponding information flow graph is updated through establishing $d$ directed edges of capacity $\beta$, starting from outgoing nodes affiliated to the selected storage nodes and terminating to the corresponding incoming node of the newcomer (See Figure~\ref{fig::modelA}). In this case, the total information received by the newcomer node, $d\beta$, is called the repair bandwidth ($\gamma$). Finally, the data is reconstructed at each DC node through connecting to any arbitrary set of $k$ nodes, including the newcommer nodes.



The use of a $[n,k,d]$ regenerating code having an access to the data of $k$ storage nodes out of existing $n$ nodes is adequate to reconstruct the original data file. Thus, the newcomer needs to connect to exactly $d = k$ nodes and downloads all of stored data ($\alpha = M/k$), thus $\beta = \alpha = M/k$. So the repair bandwidth is the same as the size of data file, i.e., $\gamma = d\beta = M$. On the other hand, Dimakis et al. in~\cite{dim10} show that if a newcomer could connect to more than $k$ surviving nodes and downloads a certain fraction of their stored information, a lower repair bandwidth would be achieved.

To this end, it is shown the task of computing the repair bandwidth can be translated to a multicast problem over the corresponding information flow graph for which an optimal trade-off between the storage per node, $\alpha$, and the repair bandwidth, $\gamma$, is identified. This optimal trade-off curve includes two extremal points corresponding to the minimum storage capacity (MSR) per node and minimum repair bandwidth (MBR), respectively.

Consider any given finite information flow graph $G$, with a finite set of data collectors. In~\cite{dim10}, it is argued that if $\min(mincut(S,DC)) \ge M$, then there exists a linear network code such that all data collectors can recover the data object.

From this symmetric model, the mincut is computed and lower bounds on the
parameters $\alpha$ and $\gamma$ are given. Let
$\alpha^{*}(d,\gamma)$ be the threshold function, which is the
function that minimizes $\alpha$.

Figure~\ref{fig::modelA} illustrates an information flow graph $G$ associated
to a $[4,2,3]$ regenerating code. Note that $mincut(S, DC) =
min(3\beta, \alpha) + min(2\beta, \alpha)$. In general, it can be claimed that $mincut(S,DC) \ge \sum_{i=0}^{k-1} min((d-i)\beta,\alpha) \ge M$, which after an optimization process leads to the threshold function shown in~\cite{dim10}.

To find the mincut equation, the $k$ terms in the summation are
computed as the minimum between two parameters: the sum
of the weights of the arcs that we have to cut to isolate the
corresponding $v^j_{in}$ from $S$, and the weight of the arc that we have to cut to isolate the corresponding $v^j_{out}$ from $S$. Let's call the first parameter as the income of the corresponding newcomer $s_j$. Note that the income of the newcomer $s_j$ depends on the previous newcomers. The newcomers can be ordered according to their income from the highest to the lowest. Then, the MSR point corresponds to the lowest income, which is given by the last newcomer added to the information flow graph; and the MBR point corresponds to the highest, which is given by the first newcomer.


\subsection{Static Cost Model} 
\label{sub:static_cost_model}

In~\cite{Akh10}, Akhlaghi et al. presented another DSS model, where
the storage nodes $V_S$ are partitioned into two sets $V^1$ and
$V^2$ with different repair bandwidth. Let $V^1 \subset V_S$ be the
``cheap bandwidth'' nodes, where each data unit has a sending cost
$C_c$, and $V^2 \subset V_S$ be the ``expensive bandwidth'' nodes,
where each data unit has a sending cost $C_e$ with $C_e > C_c$. When a newcomer enter the system, the cost of downloading data from a node in $V^1$ will be lower than the cost of downloading data from a node in $V^2$.

Consider the same situation as in the model described in
Subsection~\ref{sub:first_model}. When a storage node fails, the
newcomer node $s_j, j = n+1,\dots,\infty,$ connects to $d_1$
existing storage nodes from $V^1$ and receives from each one of them
$\beta_c$ data units; it also connects to $d_2$ existing storage
nodes from $V^2$ and receives from each one of them $\beta_e$ data
units. Let $d = d_1+d_2$ be the number of helper nodes. Assume that
$d, d_1,$ and $d_2$ are fixed, that is, they do not depend on the
storage node $s_j, j = n+1,\dots,\infty.$ In terms of the
information flow graph $G$, there is one arc from $v^i_{out}$ to
$v^j_{in}$ of weight $\beta_c$ or $\beta_e$ respectively (depending
on whether $s_i$ sends $\beta_c$ or $\beta_e$ data units) in the
regenerating process. The new vertex $v^j_{in}$, is also connected
to its associated $v^j_{out}$ with an arc of weight $\alpha$.

Let the repair cost be $C_T = d_1C_c\beta_c + d_2C_e\beta_e$ and the
repair bandwidth $\gamma = d_1\beta_c + d_2\beta_e$. To simplify the
model, we can assume, without loss of generality, that $\beta_c =
\tau \beta_e$ for some real number $\tau \ge 1$. This means that we
minimize the repair cost $C_T$ by downloading more data units from
the ``cheap bandwidth'' set of nodes $V^1$ than from the ``expensive
bandwidth'' set of nodes $V^2$. Note that if $\tau$ is increased,
the repair cost is decreased and vice-versa.


\subsection{Two-Rack Model} 
\label{sub:two_racks_model}

In~\cite{Ber13}, a new DSS model - two-rack model is presented. In
this case, the repair cost between nodes that are in the same rack
is much lower than between nodes that are in the other rack.
Consider the same situation as in
Subsection~\ref{sub:static_cost_model}, but now the sets of ``cheap
bandwidth'' and ``expensive bandwidth'' nodes are not static or
predefined, they depend on the specific replaced node.

Let the newcomers be $s_j, j=n+1,\dots,\infty$, $d^i_c$ be the
number of helper nodes providing cheap bandwidth, and $d^i_e$ be the
number of helper nodes providing expensive bandwidth to the newcomer
in the $i$-th rack, $i=1,2.$ The system must satisfy $d=
d^i_c+d^i_e$ for all $i$. Without lost of generality, assume $d^1_c
\le d^2_c$. There is a different repair bandwidth for both racks,
i.e. $\gamma^1 = \beta_e(d^1_c\tau + d^1_e) \le \gamma^2 =
\beta_e(d^2_c\tau+d^2_e)$. Recall that $\beta_c = \tau\beta_e$,
where $\tau \ge 1$. If the $\gamma^1 \ge \alpha$ is not satisfied
then the file cannot be restored.

In this model, it is not straightforward to determine which is the
set of newcomers that minimize the mincut. This set may change
according to the parameters of the system. The authors of
\cite{Ber13} show how to find the mincut set as follows: let $I$ be
the indexed multiset containing the incomes of $k$ newcomers that
minimizing the mincut.

\begin{itemize}
    \item Define $I_1 = \{((d^1_c - i)\tau + d^1_e)\beta_e | i = 0,\dots,min(d^1_c,k - 1)\}$ as the indexed multiset where $I_1[i], i = 0,\dots, min(d^1_c , k - 1),$ are the incomes of this set of $d^1_c + 1$ newcomers from rack $1$.
    \item Define $I_2 = \{d^1_e\beta_e | i = 1,\dots,min(k-d^1_c-1,n_1-d^1_c -1)\} \cup \{(d^2_c-i)\tau\beta_e | i = 0,\dots,min(d^2_c,k-n_1-1)\}$ as the indexed multiset where $I_2[i], i = 0,\dots,k -d^1_c-2,$ are the incomes of a set of $k-d^1_c-1$ newcomers, including the remaining newcomers from rack $1$ and newcomers from rack $2$.
    \item Define $I_3 = \{(d^2_c-i)\tau\beta_e | i = 0,\dots,min(d^2_c,k-d^1_c-2)\}$ as the indexed multiset where $I_3[i], i = 0,\dots,k-d^1_c-2,$ are the incomes of a set of $k-d^1_c-1$ newcomers from rack $2$.
    \item Then, either $I =I_1 \cup I_2$ or $I = I_1 \cup I_3$.
\end{itemize}

Let $L$ be the increasing ordered list of values such that for all
$i, i = 0,\dots,k-1, I[i]/\beta_e \in  L$ and $|I| = |L|$. Note that
any of the information flow graphs representing any model from this
two-rack model can be described in terms of $I$, so they can be
represented by $L$. Therefore, once $L$ is found, it is possible to
find the parameters $\alpha$ and $\beta_e$ (and then $\gamma$ or
$\gamma^i, i = 1,2$) using the following threshold function.

$$
\alpha^{*}(\beta_e) =
\left\{
    \begin{array}{ll}
        \frac{M-g(i) \beta_e}{k-i}, & \mbox{if } \beta_e \in [f(i),f(i-1)), \\
                                   & i=0,\dots,k-1,
    \end{array}
\right.
$$
subject to $\gamma^1 = (d_c^1\tau + d_e^1)\beta_e \ge \alpha,$ where
$$
f(i)=\frac{M}{L[i](k-i) + g(i)} \mbox{ and } g(i)=\sum_{j=0}^{i-1}L[j].
$$
Note that, $f(-1) = +\infty$ and $g(0)=0$ must be defined.

\section{Achieving MBR for Two-Rack Model}
\label{sec::results}

In this section, we first show that the two-rack model
in~\cite{Ber13} has an issue to achieve MBR point. A solution based
on non-homogenous distributed storage model is proposed, and then a
generalization of the threshold function is given. Finally, there is
an example comparing the traditional and non-homogeneous two-rack
models where the improvement is presented.

\subsection{Feasibility of MBR point} 
\label{sub:non_feasible_points_at_mbr} We show that in the two-rack
model presented in~\cite{Ber13} there are some situations where the
MBR point is not feasible, this is because the condition $\gamma^1 =
(d_c^1\tau + d_e^1)\beta_e \ge \alpha$ is not satisfied.

From \cite{Ber13}, the value of $\alpha$ at the MBR point is
$\alpha_{MBR} = \max(I).$ It is clear that $\max(I) = \max(I_1)$, or
$\max(I) = \max(I_2)$, or $\max(I) = \max(I_3)$, depending on the
situation. It is easy to see that $\max(I_1) = ((d^1_c - i)\tau +
d^1_e)\beta_e$ for $i = 0$, and $\max(I_1) = \gamma^1$. Hence, if
$\max(I) = \max(I_1) = \gamma^1$, then $\alpha_{MBR} = \gamma^1$,
and $\gamma^1 \ge \alpha_{MBR}$ holds.

However, if $\max(I) = \max(I_2)$ or $\max(I) = \max(I_3)$, then
$\alpha_{MBR} = \max(I) > \max(I_1) = \gamma^1$, which breaks the
required condition of $\gamma^1 \ge \alpha_{MBR}$. This implies that
some nodes receive less information than the information required
for storing during the regenerating process, and this leads to
contradiction.

The authors of~\cite{Ber13} avoid such situation by deleting as much
elements of multisets $I_2$ or $I_3$ as possible until $\max(I) =
\max(I_1)$. Such solution avoids the impossible points, but at the
same time, it also ignores better bounds in the tradeoff curve
between $\alpha$ and $\beta_e$. In other words, this is not an
efficient solution.

In fact, it is not difficult to find a case where $\max(I) = \max(I_3)$. This 
happens when $\max(I_3) > \max(I_1)$, i.e $d_c^2\tau
> d_c^1\tau + d^1_e$. For example, two in Figure~\ref{mod::c} with $\tau=3$, $d_c^1 = 1, d_e^1 = 2, d_c^2 = 2,
d_e^2 = 1$. Hence, $3\cdot2 > 1\cdot3+2$. In fact, the greater the
difference between the two racks, the greater the likelihood of this
situation will happen.

\subsection{Non-homogeneous two-rack model} 
\label{sub:introducing_a_non_homogeneous_system_to_achieve_the_mbr_point}

In this subsection we design a non-homogeneous two-rack DSS model,
and we prove that this design can achieve the MBR point that is not
feasible previously.

In the traditional two-rack model, the storage capacity of every
node is considered to be the same, say $\alpha$. Even though, the
system has two different repair bandwidths
$(\gamma^1,\gamma^2)$ for each rack. The fixed $\alpha$ and
different $\gamma$ are causing the
non-feasible points described above.

Assuming that $\gamma^2 \ge \gamma^1$, the nodes of the rack $2$ are
receiving $\gamma^2 / \gamma^1$ more information than the nodes of
rack $1$.

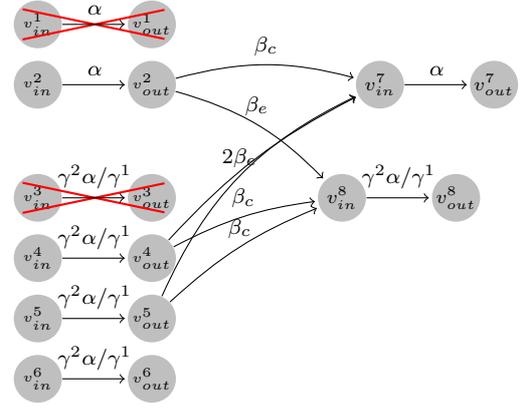
\begin{figure}
\centering
\begin{tikzpicture}[shorten >=1pt,->]
  \tikzstyle{vertix}=[circle,fill=black!25,minimum size=18pt,inner sep=0pt,
node distance = 0.8cm,font=\tiny]
  \tikzstyle{invi}=[circle]
  \tikzstyle{background}=[rectangle, fill=gray!10, inner sep=0.2cm,rounded
corners=5mm]


  \node[vertix]  (vin_1)  {$v_{in}^1$};
  \node[vertix, below of=vin_1,] (vin_2)  {$v_{in}^2$};
  \node[vertix, below of=vin_2, node distance = 1.5cm] (vin_3)  {$v_{in}^3$};
  \node[vertix, below of=vin_3] (vin_4)  {$v_{in}^4$};
  \node[vertix, below of=vin_4] (vin_5)  {$v_{in}^5$};
  \node[vertix, below of=vin_5] (vin_6)  {$v_{in}^6$};

  \node[vertix, right of=vin_1, node distance = 1.5cm] (vout_1)  {$v_{out}^1$};
  \node[vertix, right of=vin_2, node distance = 1.5cm] (vout_2)  {$v_{out}^2$};
  \node[vertix, right of=vin_3, node distance = 1.5cm] (vout_3)  {$v_{out}^3$};
  \node[vertix, right of=vin_4, node distance = 1.5cm] (vout_4)  {$v_{out}^4$};
  \node[vertix, right of=vin_5, node distance = 1.5cm] (vout_5)  {$v_{out}^5$};
  \node[vertix, right of=vin_6, node distance = 1.5cm] (vout_6)  {$v_{out}^6$};

\draw[-,color=red,thick] (-0.2,0.2) -- (1.7,-0.2);
\draw[-,color=red,thick] (-0.2,-0.2) -- (1.7,0.2);
\draw[-,color=red,thick] (-0.2,-2.1) -- (1.7,-2.5);
\draw[-,color=red,thick] (-0.2,-2.5) -- (1.7,-2.1);

  \node[vertix, right of=vout_2, node distance = 3cm] (vin_7)
{\scriptsize{$v_{in}^7$}};
  \node[vertix, right of=vin_7, node distance = 1.5cm] (vout_7)
{\scriptsize{$v_{out}^7$}};

  \node[vertix, right of=vout_3, node distance = 2.5cm] (vin_8)
{\scriptsize{$v_{in}^8$}};
  \node[vertix, right of=vin_8, node distance = 1.5cm] (vout_8)
{\scriptsize{$v_{out}^8$}};

  \path[->, bend left=15,font=\footnotesize] (vout_2) edge
node[anchor=south,above]{$\beta_c$} (vin_7);
  \path[->, bend left=10,font=\footnotesize] (vout_4) edge
node[anchor=south,above]{}(vin_7);
  \path[->, bend left=20,font=\footnotesize] (vout_5) edge
node[anchor=south,above]{$2\beta_e$}(vin_7);

  \path[->, bend left=15,font=\footnotesize] (vout_2) edge
node[anchor=south,above]{$\beta_e$} (vin_8);
  \path[->, bend left=10,font=\footnotesize] (vout_4) edge
node[anchor=south,above]{$\beta_c$}(vin_8);
  \path[->, bend left=10,font=\footnotesize] (vout_5) edge
node[anchor=south,above]{$\beta_c$}(vin_8);

 \foreach \from/\to in {1,2,7}
  { \path[->,font=\footnotesize] (vin_\from) edge node[anchor=south] {$\alpha$}
(vout_\to); }
 \foreach \from/\to in {3,4,5,6,8}
  { \path[->,font=\footnotesize] (vin_\from) edge node[anchor=south] {$\gamma^2\alpha/\gamma^1$}
(vout_\to); }

 \end{tikzpicture}

\caption{Non-homogeneous two-rack DSS model. Rack $1$ with two nodes and rack $2$ with four nodes. Note that, $\gamma^1 = \beta_c + 2\beta_e$ and $\gamma^2 = 2\beta_c + \beta_e$.}
\label{fig::nonhomo}
\end{figure}

Our approach is to design a non-homogeneous two-rack model where the
nodes of rack $1$ stores $\alpha$ information and the nodes of rack
$2$ stores $\frac{\gamma^2}{\gamma^1}\alpha$ information. Recall
$\gamma^1 = \beta_e(d^1_c\tau + d^1_e) \le \gamma^2 =
\beta_e(d^2_c\tau+d^2_e)$. Figure~\ref{fig::nonhomo} shows this new
model.


In the proposed non-homogeneous two-rack model, the mincut equation,
which is not constant in terms of $\alpha$ (as it was in the
original two-rack model), becomes:

$$
C = \min{\{I[i],\alpha\}} + \min{\{I[j],\frac{\gamma^2}{\gamma^1}\alpha\}},
$$
where $I[i]$ are the incomes of the rack $1$, and $I[j]$ are the
incomes of the rack $2$.

Note that, the mincut set for the newly proposed non-homogeneous
two-rack model is still the same as the traditional two-rack model.
Hence, the set of incomes $I$ is exactly the same. The main
difference arises in $L$. In the traditional two-rack model, the
list $L$ is created in ascendant order by picking the elements of
$I$. Let's define the following multiset of tuples:
$$
L^{n} = \{(\frac{I[i]}{\beta_e},1)\} \cup \{(\frac{I[j]}{\beta_e},\frac{\gamma^2}{\gamma^1})\}
$$
where $I[i]$ are the incomes of the rack $1$ and $I[j]$ are the
incomes of the rack $2$. Moreover, $L^{n}$ is ordered by the
following total order:
$$ L^{n}[i] \ge L^{n}[j] \Longleftrightarrow L^{n}[i][1]L^{n}[i][2]^{-1} \ge L^{n}[j][1]L^{n}[j][2]^{-1}.$$
Next, we can generalize the threshold function for the
non-homogeneous two-rack model:
$$
\alpha^{*}(\beta_e) =
\left\{
    \begin{array}{ll}
        \frac{M-g'(0,i,1) \beta_e}{g'(i,k-1,2)}, & \mbox{if } \beta_e \in [f(i),f(i-1)),
    \end{array}
\right.
$$
for $i=0,\dots,k-1,$ where
$$
f(i)=\frac{M}{g'(0,i,1)+g'(i,k-1,2)L^n[i][1]L^n[i][2]^{-1}}
$$
and
$$
g'(a,b,c)=\sum_{j=a}^{b}L^n[j][c].
$$
Note that, $f(-1) = +\infty$.

The next theorem shows how all the points on the tradeoff curve are
feasible in the newly proposed non-homogeneous two-rack model.

\begin{theo}
    Given a non-homogeneous two-rack model with repair bandwidths $\gamma^1 \le \gamma^2$ and the nodes of the rack $1$ stores $\alpha$ information and the nodes of rack $2$ stores $\frac{\gamma^2}{\gamma^1}\alpha$ information. Then, all the points of the tradeoff curve are feasible.
\end{theo}
\begin{IEEEproof}
    As in the traditional two-rack model, $\alpha_{MBR}$ is defined by the maximum income. But now, the ``maximum income'' is taken from the multiset $L^n$ and depending on the total order defined above (definitely it depends on the storage too). Thus, we need to prove that the ``maximum income'' is always $\gamma^1$. The problem can be translated to the multiset $I$. Since it is constructed by $I_1,I_2$ from the rack $1$ and $\frac{\gamma^1}{\gamma^2} I_3$ from the rack $2$, we need to show that $\alpha_{MBR} = \max(I) = \max(I_1) = \gamma^1.$

   Since $\max(I) = \max(I_1 \cup I_2)$ or $\max(I) = \max(I_1 \cup \frac{\gamma^1}{\gamma^2}I_3)$. And $\max(I_1) = \gamma^1 = (d^1_c\tau + d^1_e)\beta_e$, $\max(I_2) = d_e^1\beta_e \le \gamma^1$. Then, in this case, $\max(I) = \max(I_1 \cup I_2) = \max(I_1) = \gamma^1$. On the other hand, if we consider $\frac{\gamma^1}{\gamma^2} I_3$, we can see that $\max(I_3) = d_c^2\tau\beta_e \le \gamma^2 = d_c^2\tau\beta_e + d_e\beta_e$. Hence, $\max(\frac{\gamma^1}{\gamma^2} I_3) = \gamma^1\frac{d_c^2\tau\beta_e}{\gamma^2} \le \gamma^1$. And it holds too that $\max(I_1 \cup \frac{\gamma^1}{\gamma^2} I_3) = \max(I_1) = \gamma^1$

    Finally, the MBR point becomes feasible without the need of deleting any element of the list $I$. Since $\gamma^1 \ge \alpha$ then $\gamma^2 \ge \frac{\gamma^2}{\gamma^1}\alpha$.
\end{IEEEproof}

\subsection{Example} 
\label{sub:example} 
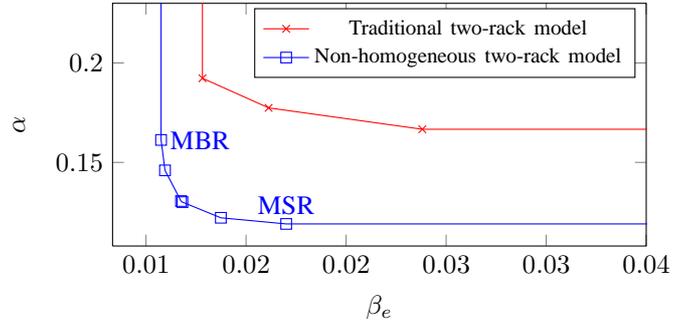
\begin{figure}
\centering

\begin{tikzpicture}
 \begin{axis}
[ xlabel=$\beta_e$,
  ylabel=$\alpha$,
  xmax= 0.035,
  ymax=0.23,
  height=4.8cm,
  width=8.6cm,
  scaled x ticks = false,
  legend style={font=\footnotesize},
  x tick label style={/pgf/number format/fixed, /pgf/number format/1000 sep =
\thinspace}
]

\addplot[color=red,mark=x] coordinates { 
(1.1, 0.166666666666666666666666666667)
(0.0238095238095238095238095238095,0.166666666666666666666666666667)
(0.0161290322580645161290322580645,0.177419354838709677419354838710)
(0.0128205128205128205128205128205,0.192307692307692307692307692308)
(0.0128205128205128205128205128205,1)};

\addplot[
    color=blue,
    mark= square,
    visualization depends on=\thisrow{alignment} \as \alignment,
    nodes near coords, 
    point meta=explicit symbolic, 
    every node near coord/.style={anchor=\alignment} 
    ] table [
     meta index=2 
     ] {
x                                 y                                   label       alignment
1.1000000000000000000000000000000 0.119047619047619047619047619048    \,           0
0.0170068027210884353741496598639 0.119047619047619047619047619048    MSR          270
0.0137404580152671755725190839695 0.122137404580152671755725190840    \,           0
0.0118203309692671394799054373522 0.130023640661938534278959810875    \,           0 
0.0117493472584856396866840731070 0.130548302872062663185378590079    \,           0 
0.0109489051094890510948905109489 0.145985401459854014598540145985    \,           0
0.0107526881720430107526881720430 0.161290322580645161290322580645    MBR          180
0.0107526881720430107526881720430 1.000000000000000000000000000000    \,           0
};
\legend{Traditional two-rack model, Non-homogeneous two-rack model}
\end{axis}
\end{tikzpicture}

\caption{Chart comparing the traditional and the non-homogeneous two-rack models. $M=1, k=6, d^1_e=7, d^2_e=3, d=9, n_1=3,n_2=7, \tau=4$.}
 \label{fig::grafica}
\end{figure}

A comparison between the traditional two-rack
model and the newly proposed non-homogeneous two-rack model is shown
in Figure~\ref{fig::grafica}. We consider a two-rack model with $3$
nodes in the first rack,  $7$ nodes in the second rack, and with
$\tau = 4$. Three points has been deleted in the traditional model.
The non-homogeneous case not only achieves the MBR, the performance
on the MSR is also better, even this is not due to any deleted point
in the traditional model.

\section{Conclusion}
\label{sec::conc} In this paper, we show that a traditional two-rack
DSS model that considering only different repair bandwidth across the rack
but same storage size for all the nodes cannot achieve the MBR
point. We propose a non-homogenous model by having a different
storage size for the storage nodes in each rack, and prove that this
non-homogenous model makes MBR point becomes feasible. Moreover, we
show how much information should be stored on each node and derive a
generalized threshold function. The generalization of this
non-homogeneous model to any number of racks is straightforward after the traditional two-rack model is generalized.

\section{Acknowledgment}
This research is partly supported by the International Design Center (grant no. IDG31100102 and IDD11100101), the Spanish MICINN grant TIN2010-17358, the Spanish Ministerio de Educación FPU grant AP2009-4729 and the Catalan AGAUR grant 2009SGR1224.



\begin{thebibliography}{1}

%

\bibitem{rod05}
  R. Rodriguez, B. Liskov, ``High availability in dhts: Erasure coding vs. replication'' in \emph{ Proceedings of the IPTPS05}. 2005.

\bibitem{wea02}
  H. Weatherspoon, J. Kubiatowicz, ``Erasure coding vs replication: a quantitative comparison'' in \emph{ Proceedings of International Peer-to-Peer Systems}, vol 2429, pp 328--337 2002.

\bibitem{dim10}
  A. Dimakis, P. Godfrey, M. Wainwright, K. Ramchandran, ``Network Coding for Distributed Storage Systems'' in \emph{IEEE Trans. on Inf. Theory}, vol 59 no. 9, pp. 4539--4551, 2010.

\bibitem{ras11}
  K. V. Rashmi, Nihar B. Shah, and P. Vijay Kumar, ``Optimal Exact-Regenerating Codes for Distributed Storage at the MSR and MBR Points via a Product-Matrix Construction'' in \emph{IEEE Trans. on Inf. Theory}, vol 57 no. 8, pp. 5227--5239, 2011.

\bibitem{rou10}
  S. El Rouayheb, K. Ramchandran, ``Fractional repetition codes for repair in distributed storage systems'' in \emph{48th Annual Allerton Conference on Communication, Control, and Computing}, 1510--1517, 2010.

\bibitem{Akh10}
    S. Akhlaghi, A. Kiani, M. Ghanavati, ``A fundamental trade-off between the download cost and repair bandwidth in distributed storage systems,'' IEEE Int. Symp. on Network Coding NetCod, pp. 1–6, 2010.

\bibitem{Van12}
    V. T. Van, C. Yuen, J. Li,``Non-homogeneous distributed storage systems,'' in Allerton, 2012.

\bibitem{Ern12}
    T. Ernvall, S. E. Rouayheb, C. Hollanti, V. Poor ``Capacity and Security of Heterogeneous Distributed Storage Systems,'' in arXiv:1211.0415v1, 2012.

\bibitem{Ber13}
    B. Gast\'on, J. Pujol, M. Villanueva ``A realistic distributed storage system that minimizes data storage and repair bandwidth'', Data Compression Conference 2013. preprint at http://arxiv.org/abs/1301.1549


\end{thebibliography}
\end{document}